\documentclass[aps,prb,twocolumn,showpacs,preprintnumbers,amsmath,amssymb,floatfix]{revtex4-2}
\usepackage{graphicx}
\usepackage{dcolumn}
\usepackage{bm}

\usepackage{epstopdf}%
\usepackage{amsmath}%
\usepackage{amsfonts}%
\usepackage{amssymb}
\usepackage{xcolor}
\definecolor{Mycolor2}{HTML}{FF00FF
}

\newcommand{\red}{\textcolor{black}}
\newcommand{\green}{\textcolor{black}}
\newcommand{\blue}{\textcolor{black}}
\newcommand{\mycolor}{\textcolor{black}}

\begin{document}

\title
{Conductance zeros in complex molecules and lattices from the interference \green{set} method}

\author{M. Ni\c t\u a}
\author{M. \c Tolea}
\affiliation{National Institute of Materials Physics, Atomistilor 405A,
Magurele 077125, Romania}
\author{D. C. Marinescu}
\affiliation{Department of Physics, Clemson University, Clemson, SC 29634}


\begin{abstract}
%
Destructive quantum interference (DQI) and its effects on electron transport is studied in chemical molecules and finite physical lattices that can be described by a discrete Hamiltonian. Starting from a bipartite system whose conductance zeros are known to exist between any two points of a specially designated set, the interference set, we \green{use the Dyson equation to} develop a general algorithm of determining the zero conductance points in complex systems, which are not necessarily bipartite. We illustrate this procedure as it applies to the fulvene molecule. The stability of the conductance zeros is analyzed in respect with external perturbations.

\end{abstract}

\date{\today}
\pacs{72.80.Vp,73.21.La, 73.63.Kv}
\maketitle

\section{Introduction}


Quantum interference effects associated with the electron propagation in a conductor have long been a subject of interest in mesoscopic physics.
Depending on the phase difference between the different electron paths the conductance can fluctuate as in
the Aharonov-Bohm effect \cite{aharonov1959, oded2008, imry2008} or
completely vanish in various phase coherent systems \cite{lee1999, julian2009, sparks2011, liang2012,
lambert2015, pedersen2015, tsuji2016, timoti2016,
zhao2017, yanxi2018, junyang2018, lambert2018, zainelabideen2019,
naxin2019, yueqi2019, li2019}.
The latter situation, associated with a destructive quantum intereference (DQI),
is equivalent to a zero propagation probability for the electron state
sometimes referred to as ``antiresonance'' \cite{gunasekaran2020, yangli2019}. In electron transport, within the Landauer-B\"{u}ttiker formalism, the existence of a DQI between two lattice sites is equivalent to a conductance zero between
the same sites \cite{levy2000, nita2020}.

The DQI has been studied in different physical and chemical systems, such as quantum dots \cite{levy2000, rotter2005},
graphene-type structures \cite{tada2002, nita2014, valli2019},
benzene and other carbon based molecules \cite{sautet1988, nozaki2017, sykora2017a, sykora2017b, nozaki2017b, pedersen2014, tsuji2014, shuguang2018}
and T-shape conductors \cite{andrey2010, daijiro2013}.
Several theoretical methods have been used to determine if zero conductance points appear in a non-interacting lattice, such as
the wave functions parity method \cite{levy2000},
graph based selection rules \cite{fowler2009, mayou2013},
the unpaired atoms graphical method \cite{markussen2010}. Other approaches involved
the linear algebra of the molecular Hamiltonian in the presence of electrode couplings \cite{matthew2014},
curly arrows \cite{stuyver2015},
interference vectors \cite{sam2017},
the calculation of the conductance cancellations using the
characteristic polynomials \cite{tsuji2018}
or the identification of the bipartite sublattice blocks with well known conductance zeros  \cite{nita2020}.
\red {A general perspective on selection rules for destructive quantum interference in single-molecule
electron transport was given in Ref.~\cite{evers2020}}.

In this paper
\green{we develop an algorithm based on the Dyson equation} to determine the existence of DQIs between pairs of sites in a non-interacting electron system described by a discrete Hamiltonian (H\"{u}ckel or tight-binding).
\mycolor{First, we consider a bipartite system and use its symmetry properties to determine the pairs of lattice points between which DQI occurs. The ensemble of these sites defines \green{the interference set},
${\cal M}_I$.}
Although ${\cal M}_I$ can be established for any electron energy E,
here we focus on the mid-spectrum propagation modes
since in bipartite lattices they correspond to well-defined classes of conductance zeros - such as those in half-filled graphene \cite{tada2002, nita2020}. The configuration of the ${ \cal M}_I$ set is discussed in Section\,II.
This preamble is then used in Section\,III to formulate the general conditions that assure the existence of DQI points in complex systems, not necessarily bipartite. The invariance of the conductance zeros under several different perturbations is analyzed. We illustrate the application of this algorithm to fulvene, a non bipartite molecule in Section\,IV. Our conclusions are stated in Section\, V.

\green{\section {The interference set of a bipartite lattice}}

\begin{figure}[]
 \includegraphics[scale=1.0]{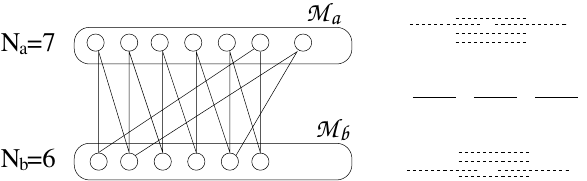}\nonumber
\caption {A generic bipartite system
with $N_a=7$ and $N_b=6$. The inter-sublattice hopping are leading to zero energy states with total degeneracy $g=3$.
The sketch of the energy spectrum is on the right.}
\label{reteabip}
\end{figure}

A bipartite system consists of two sublattices $A$ and $B$ whose sets of points $\mathcal M_a$ and $\mathcal M_b$, are coupled through hopping matrix elements, $h_{n_a,n_b} \neq 0$, with $n_a\in {\mathcal M_a}$ and $n_b\in {\mathcal{M}_b}$, as described in
 Fig.\,\ref{reteabip}. The Hamiltonian is written as,
\begin{eqnarray}\label{hbip}
H_{bip}=\sum_{n_a,n_b} h_{n_a,n_b} | n_a \rangle \langle n_b |+ \text{H.c.}\,, 
\end{eqnarray}
where, for simplicity, the energies $h_{n_a,n_b}$ are usually assumed to be all equal (and the energy unit, i.e. $t = 1$).
Eq.~(\ref{hbip}) can describe
chemical molecules \cite{fowler2009, stuyver2015},
nanostructures \cite{tada2002, dhakal2019}
or artificial molecules composed of quantum dots \cite{tamura2002, norimoto2018}.

$H_{bip}$ anticommutes with the chirality operator $\Gamma$,
\begin{eqnarray}\label{gamma}
\Gamma=\sum_{n_a} | n_a \rangle \langle n_a |-\sum_{n_b}| n_b \rangle \langle n_b |\,,  
\end{eqnarray}
thus ensuring that the eigenstates of the system satisfy $\Psi_{- \epsilon}=\Gamma \Psi_{ \epsilon}$. Consequently,
\begin{eqnarray}\label{pair}
\Psi_{\pm \epsilon} =\sum_{n_a} \Psi ({n_a}) | n_a \rangle \pm
          \sum_{n_b} \Psi ({n_b}) | n_b \rangle\,, 
\end{eqnarray}
a property related to the pairing theorem \cite{coulson1940} or the electron-hole symmetry \cite{deng2014, ostahie2016, nita2017}.

The degeneracy $g$ of the eigenvalue $\epsilon = 0$ is determined by the expansion of the determinant of $H_{bip}$.
For example, in Fig.\,\ref{reteabip} we show a bipartite lattice and its energy spectrum which in that case corresponds to $g=3$.
If $H_{bip}$ is singular, $\det H_{bip} = 0$, $g\ne 0$, and there are one or more zero energy eigenstates $\Psi_{0i}$ with
$i=1,\cdots,g$.
For a non-singular $H_{bip}$, $\det H_{bip} \ne 0$ and there is no $\epsilon=0$ eigenstate.
The necessary (but not sufficient) condition for this later situation to occur in a bipartite lattice is $N_a=N_b$
\cite{koshino2014}.
%

%

Using the energy eigenstates and eigenvectors, $\epsilon_\alpha$ and $\Psi_\alpha$ with $\alpha = 1,\cdots, N$,
we calculate the matrix elements of the
the Green's function operator at a given energy $E$,
\begin{eqnarray}
G(E)=\frac{1}{E-H_{bip}}\;.
\end{eqnarray}
For propagation between points in the same sublattice (say, sublattice "A") we obtain, making use of Eq.~(\ref{pair}),
\begin{eqnarray}\label{gaa}
G_{n_am_a}(E)&=& \sum_{\alpha, \epsilon_\alpha>0} \frac {2E}{E^2-\epsilon _\alpha^2} \Psi_\alpha (n_a) \Psi_\alpha^* (m_a)\nonumber\\
&+& \sum_{i=1}^{g}\frac{\Psi_{0i} (n_a) \Psi_{0i}^* (m_a)}{E}\,.
\end{eqnarray}
%
In a non-singular bipartite system, with $g = 0$ and no energy levels at mid-spectrum, Eq.\,\ref{gaa} indicates that,
$G_{n_am_a}(0) = 0$. Therefore DQIs occur between any of the $A$ points at $E=0$ as previously discussed
in the context of conductance cancellations in Refs.~\citenum{levy2000, tsuji2018, nita2014, nita2020, evers2020}.
\red {By using the chirality property from Eq.\,\ref{pair},
the terms with opposite energies from the spectral decomposition of the Green's function in Eq.\,\ref{gaa}
cancel each other. This is in agreement with the fact that
the ${G}_{AA}$  zeroes can also be  explained
as destructive interference in the energy space \cite{nozaki2017, gunasekaran2020}.}
In the weak coupling limit this is mostly reduced to the tunneling through the
\green{ two adjacent energy states \cite{levy2000},
HOMO and LUMO in molecules \cite{tsuji2017}.}


The points of a molecule/lattice between which
the matrix element of the Green's function cancels at a given energy,
say $G_{nm}(E=0) = 0$, form what we define as an "interference set", $\mathcal{M}_I$.
Thus,
\begin{eqnarray}\label{gfip}
G_{nm}(E)=0,~~\forall~ n,m\in {\mathcal M}_{I}\,.~
\end{eqnarray}
The case $n=m$ is necessary to be included since $G_{nn}(E)=0$ is fulfilled for any $n\in {\mathcal M}_{I}$.
 Within the Landauer-B\"{u}ttiker formalism the cancellation of the bare Green's functions determines the cancellation of the electrical conductance.
 This is obtained by using Eqs.\,\ref{geff}-\ref{tra} evaluated at a Fermi energy equal to $E$ (see Appendix A).
\green{Therefore,}
\begin{eqnarray}\label{gfip2}
{\bf G}_{nm}(E)=0,~\forall~ n,m\in {\mathcal M}_{I}\,,
\end{eqnarray}
\green{where $n$ and $m$ are the external lead contact points, i.e. source and drain.}

In a bipartite lattice a set ${\mathcal M}_{I}$ is easily found at $E=0$ by using the results from Eq.\,\ref{gaa}.
For a non-singular system ($g=0$ and no $\epsilon=0$ state) one identifies two disjoint sets,
composed of all $A$ and all $B$ sublattice points, respectively, since
$G_{AA}(0)=0$ and $G_{BB}(0)=0$. So from the definition (\ref{gfip}) one obtains two interference sets:
\begin{eqnarray}\label{p11fig}
{\mathcal M}_{I}^{(1)}={\mathcal M}_a~~\text{and}~~ {\mathcal M}_{I}^{(2)}={\mathcal M}_b\,.
\end{eqnarray}


If the bipartite Hamiltonian is singular, such that $\epsilon = 0$ is an eigenvalue,
a more subtle analysis is required to determine the interference set
since in general not all matrix elements of $G_{AA}$ and $G_{BB}$ cancel.
\red{In this case, one calculates the zero energy states $\Psi_{0i}$,
that enter in Eq.\,\ref{gaa},
to see which of the $A$ (or $B$) points can be selected in the ${\mathcal M}_{I}$ set.}

In addition to the interference sets that contain only one type of points, $A$ or $B$,
mixed $I$ sets containing both type of points $A$ and $B$
can be obtained in the case of composite bipartite systems,
i.e. two serial coupled bipartite sublattices \cite{nita2020}.

\begin{figure}[ht]
\includegraphics[scale=1.1]{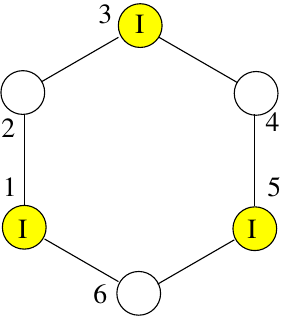}
\caption{ The hexagon lattice of the benzene molecule described by a bipartite Hamiltonian
has two equivalent sets of interference points $\mathcal M_{I} = \{1,3,5\}$ (yellow) and $\mathcal M_I = \{2,4,6\}$.
}
\label{hex-fig}
\end{figure}

To illustrate the application of these definitions, we find the interference points in the hexagon system,
related to the benzene molecule
 - which is one of the common examples of DQI in single-molecule electron transport
\cite {sautet1988, patoux1997, nozaki2017, sykora2017a, nozaki2017b, sam2017}.
As shown in Fig.~\ref{hex-fig}, this is a bipartite system whose sub-lattice sets are ${\mathcal M}_a=\{ 1, 3, 5 \}$
and ${\mathcal M}_b=\{ 2, 4, 6 \}$. Since the Hamiltonian matrix is non-singular,
from Eq.\,\ref{p11fig} the two interference sets are ${\mathcal M}^{(1)}_{I}={\mathcal M}_a$
and ${\mathcal M}^{(2)}_{I}={\mathcal M}_b$.

For ${\mathcal M}_{I}={\mathcal M}_a$ we have
$
G_{n,m}(0)=0,~(\forall)~n,m\in \{ 1, 3, 5 \}\,, 
$
while for  ${\mathcal M}_{I}={\mathcal M}_b$,
$G_{n,m}(0)=0,~(\forall)~n,m\in \{ 2, 4, 6 \}\,.$
The conductance formula from Eq.\,\ref{gfip2} determines the corresponding
transport cancellations.


\section{General Algorithm}

We start from a bipartite lattice with a known interference set $\mathcal{M}_I$.
The complete set of lattice points $\mathcal{M}$ is therefore decomposed in the reunion of two disjoint subsets,
\begin{eqnarray}\label{desc1}
{\mathcal M}={\mathcal M}_{I}\cup {\mathcal M}_{R}\,.
\end{eqnarray}
where $\mathcal{M}_R = \mathcal{M} - \mathcal{M}_I$ is by definition the rigid set.

\begin{figure}[h]
\includegraphics[scale=1.33]{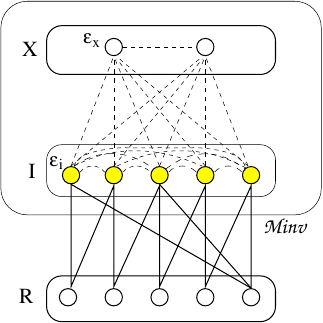}
\caption{
A complex system is built by
perturbing the $I$ points from the set ${\mathcal M}_{I}$
or by
connecting  external points $X$ from the set ${\mathcal M}_{X}$
to the interference points $I$.
The $R$ points are not affected during the construction of the new system.
The larger contour enclosing the two sets ${\mathcal M}_{X}$ and ${\mathcal M}_{I}$
defines the invariance set ${\mathcal M}_{inv}$.
}
\label{mpi1}
\end{figure}

This situation is illustrated in Fig.\,\ref{mpi1}
where \green{the ${\mathcal M}_{a}$ sublattice is chosen to be the interference $I$ set (yellow circles) and the remaining ${\mathcal M}_b$ sublattice is chosen as the rigid $R$ set (empty circles).}
\blue{While it is not mathematically necessary that the $I$ set is identical to $\mathcal M_a$, such a choice is indicated if one wants to maximize the number of conductance cancelations that are found}.
The full lines between $I$ and $R$ points correspond to the nonzero hopping energies
of the bipartite Hamiltonian.

\red {For the selected $I$ and $R$ lattice points},
the Hamiltonian of the bipartite lattice is consequently re-expressed as,
\begin{eqnarray}\label{hinput}
H = H(I,R)\,. 
\end{eqnarray}
The Green's function matrix for the ${\mathcal M_I}$ subspace states $G_{II}$ is zero,
\begin{eqnarray}\label{gii}
G_{II}(E)=0\,,
\end{eqnarray}
in agreement with Eq.\,\ref{gfip}.
$H$ is perturbed by two additional Hamiltonians, $H_1(I)$ which describes the non-zero hopping probabilities among interference points (including onsite new energy terms) and $H_2(I,X)$
that contains the hopping elements between the interference points and
a set of external points ${\mathcal M}_X$, \green{as well as any additional $X$-set related terms}.
The new terms correspond to the dashed lines or to
\red{the on-site energies such as $\epsilon_i$ and $ \epsilon_x$} in Fig.\,\ref{mpi1}.
 The resulting Hamiltonian $H'$ is
\begin{eqnarray}\label{trampi}
H'(I,R,X)=H(I,R)+H_1(I)+H_2(I,X)\,. 
\end{eqnarray}
The associated lattice $\mathcal{M'}$ described by the Hamiltonian $H'$ is the reunion
 of the three disjoint subsets,
\begin{eqnarray}\label{desc2}
{\mathcal M}'={\mathcal M}_{I} \cup {\mathcal M}_{R} \cup {\mathcal M}_{X}\,, 
\end{eqnarray}
as illustrated in  Fig.\,\ref{mpi1}.
\red{The original lattice can be grown by adding new points X, such that a conductance cancellation occurs between any point in the $X$ set and any point in the $I$ set. Thus, the conductance cancellations in the enhanced lattice are known apriori by contruction, once the interference set is established.}

We determine the DQI processes in the complex system $H'(I,R,X)$ by evaluating the matrix elements of the Green's
function operator $G'$. \red{It is assumed that they have no singularities at energy $E$.}
The Dyson equations satisfied by
the matrices $G'_{II}$ and $G'_{IX}$ are written in terms of the matrix $G_{II}$ as,
\begin{eqnarray}\label{gprim0}
&& G'_{II}= G_{II}+G_{II} h_{II} G'_{II}+ G_{II} h_{IX} G'_{XI}\;,
\\
\label{gprimext}
&& G'_{IX}=  G_{II} h_{II} G'_{IX}
            +G_{II}  h_{IX} G'_{XX}\,,~
\end{eqnarray}
where $h_{II}$ is the matrix that contains the hopping energies between $I$ sites given by
Hamiltonian $H_1(I)$ from Eq.\,\ref{trampi} and
 $h_{IX}$ is the matrix that contains the hopping energies between $I$ and $X$ points
introduced by the Hamiltonian $H_2(I,X)$. We note that the unperturbed Green's function $G_{IX}$ is zero since the $\mathcal M_I$ and $\mathcal M_X$ sets are initially uncoupled.

With $G_{II}(E)=0$ from Eq.\,\ref{gii}, Eqs.\,\ref{gprim0} and \ref{gprimext} generate,
\begin{eqnarray}\label{gprim}
&& G'_{II}(E)=0\,, ~~ 
\\
\label{gprim2}
&& G'_{IX}(E)=0\,. ~~   
\end{eqnarray}

Eq.\,\ref{gprim} indicates that ${\mathcal M}_{I}$ is a proper interference set for $H'$
in agreement with the definition from (\ref{gfip}).
All the DQI process between the $I$ points are common to both systems described by $H$ and $H'$,
as shown in Eqs.\,\ref{gii} and \ref{gprim}.
The supplementary cancellations in Eq.\,\ref{gprim2} exhibit
the appearance of new DQI processes in the output system between $I$ and $X$ points
of the lattice.

The stability of the DQIs is studied with the invariance set,
\begin{eqnarray}\label{defminv}
{\mathcal M}_{\text {inv}}={\mathcal M}_{I} \cup {\mathcal M}_{X}\,.~
\end{eqnarray}
The interference points $I$ and the Green's function zeros, ${G'}_{II}$ and ${G'}_{IX}$,
are not changed by any deformation of $H'$ related to ${\mathcal M}_{\text {inv}}$ points,
\begin{eqnarray}\label{trainv}
H'\to H'+ \sum _{n,m\in {\mathcal M}_{\text {inv}}} h_{nm} |n \rangle \langle m |\,. ~~ 
\end{eqnarray}
In Fig.\,\ref{mpi1} the transformation (\ref{trainv}) could represent
the modification of hopping energies related to the dashed lines,
to the modification of the on-site energies $\epsilon_i$ of $I$ points
or $\epsilon_x$ of $X$ sites.
$H'$ becomes an effective Hamiltonian when non-hermitian terms,
that are introduced to simulate the presence of the external leads attached to the $I$ or $X$ points,
are added \cite{tolea2010, ostahie2016, cardamone2006}.
In the figure, the ${\mathcal M}_{\text {inv}}$ set is shown by the larger contour enclosing $I$ and $X$ points.

 From the Green's functions zeros (\ref{gprim}, \ref{gprim2}) for the new molecule or physical system
we are able to easily predict its conductance zeros
when the transport measurements are performed.
By connecting the external leads to sites in ${\mathcal M}_{I}$, from Eq.\,\ref{gprim}
one obtains the conductance zeros ${\bf G}_{II}=0$.
\red {When the leads are connected to a pair of points, one from ${\mathcal M}_I$ and the other from $\mathcal{M}_X$,
 from Eq.\,\ref{gprim2} one obtains ${\bf G}_{IX}=0$}.
Any system modification according to transformation (\ref{trainv})
\red{leaves these conductance zeroes unchanged.}

The algorithm discussed here can be applied to various physical or chemical systems
that contain embedded sub-systems with known interference sets which undergo
 the decomposition from Eq.\,\ref{trampi}
or Fig.\,\ref{mpi1}.

We discuss now the case of a bipartite system for which it is {\it a priory} known that  $\epsilon = 0$ is not an eigenvalue.
For such systems
\red{any} of the sub-lattices $A$ or $B$  \red{can be selected as the ${\cal M}_I$ set of points}
with Green's function (and conductance) cancellations for any pair of sites between them,
including pair of identical sites, as discussed in Section\,II.
Such examples are
the circular systems with an even number of atoms which is not a multiple of four
and the linear chains with an even number of sites.
For the circular molecules we remind that those with $4N$ atoms have a pair of degenerate levels
 at $\epsilon =0$, which are absent in all other cases \cite{nita2017}.
The chain property is trivial \cite{tsuji2014}.

For the hexagonal system in Fig.\,\ref{hex-fig}, the general algorithm described above starts by identifying the interference sets.
According to Eqs.\,\ref{p11fig}
this bipartite system has two interference sets
${\mathcal M}_{I}^{(1)}=\{1, 3, 5\}$ and ${\mathcal M}_{I}^{(2)}=\{2, 4, 6\}$.
The corresponding rigid points sets from Eq.\,\ref{desc1} are given by ${\mathcal M}-{\mathcal M}_{I}$.
The two sets are disjointed, so if one is picked as the interference set, by default the other one becomes the rigid set.
For ${\mathcal M}_I^{(1)} = \{1,3,5\}$, ${\mathcal M}_R^{(1)} = \{2,4,6\}$ and
the invariance set from Eq.\,\ref{defminv} is obtained to be ${\mathcal M}_{\text {inv}}^{(1)}=\{1, 3, 5\}$
since in this case no external sites were added.
Following our theory, diagonal or hopping energies can be added to the ${\mathcal M}_{\text {inv}}^{(1)}$ sites
without destroying the $I$ points.

The conductance cancellations follow from the $G_{II}$ zeros of the Green's function.
For ${\mathcal M}_{I}^{(1)}=\{1, 3, 5\}$ these conductance zeros are ${\bf G}_{13}, {\bf G}_{15}$ and ${\bf G}_{35}$
for the meta-contacted benzene \cite{nozaki2017, fowler2009},
and ${\bf G}_{11}=0, {\bf G}_{33}=0$ and ${\bf G}_{55}=0$ when the two leads are connected to the same point
as discussed for the ipso-contacted benzene in \cite{fowler2009}.
All of these zeros have
the invariance set
${\mathcal M}_{\text {inv}}^{(1)}=\{1, 3, 5\}$.
For instance the invariance of ${\bf G}_{15}$ at the site $3$ perturbation
proves the robustness of the meta-contacted benzene when
 a heteroatom substitution is performed for real value of $\epsilon_3$ \cite{sara2018}
or when a third external lead is attached, in this case $\epsilon_3$ energy having a complex value \cite{cardamone2006}.
On this line our results may be a support to understand the conductance invariance
when heteroatom substitutions in some molecules are performed
\cite{garner2016, tsuji2017, tsuji2017b, liu2017, sara2018}.

\red{We stress that not all lattices have interference sets,
 even if they may have conductance cancellations
(i.e. even if the conductance between $n$ and $m$ is zero, they do not form an interference set unless the conductance is also zero when both leads are connected to $n$ and also when both leads are connected to $m$).
     The property that a system has an interference set is therefore not trivial, but rather an exception met for instance in bipartite lattices.}

\section{Extension of the formalism to non-bipartite lattices. The fulvene molecule}

In this section we present a complete analysis of the conductance zeros in fulvene, thus augmenting the results of Ref. \citenum{tsuji2018} which studied only several such instances.



\begin{figure}[h]
\includegraphics[]{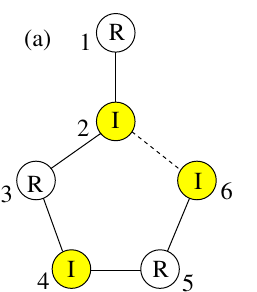}
\includegraphics[]{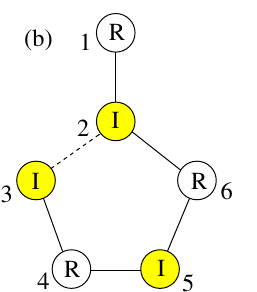}
\includegraphics[]{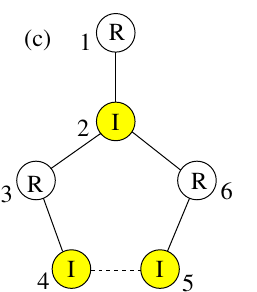}

\includegraphics[]{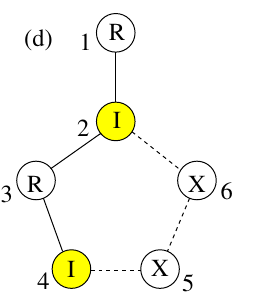}
\includegraphics[]{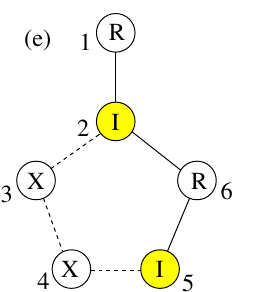}
\includegraphics[]{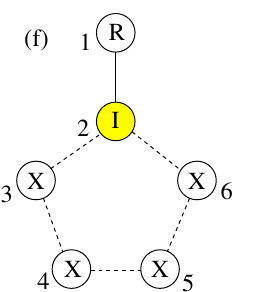}

\caption{Interference points in fulvene molecule
and the decomposition of its lattice sites ${\mathcal M}={\mathcal M}_{I}\cup {\mathcal M}_{R} \cup {\mathcal M}_{X}$.
$I\in {\mathcal M}_{I}$ and $R \in {\mathcal M}_{R}$
are the interference and rigid points
of the starting bipartite molecule $H_{bip}$ with the lattice points set
${\mathcal M}_{bip}={\mathcal M}_{I}\cup {\mathcal M}_{R}$.
In (a)-(f) the dashed lines correspond to the new hopping energies,
while the $X$ circles in (d)-(f) designate the new external sites ${\mathcal M}_{X}$
added to the initial bipartite molecule (composed of $I$ and $R$ circles). 
}
\label{fulvena}
\end{figure}

To apply the algorithm described in the previous section, we first determine the interference points sets ${\mathcal M}_{I}$, as shown in Fig.~\ref{fulvena}.
First, we determine a smaller bipartite system with Hamiltonian $H$ that can provide a set of $I$ points, such that
$H=H(I,R)$ is the input Hamiltonian in Eq.\,\ref{hinput}.
Then, the remaining sites \red{and hoppings} are added to the $I$ points, generating $H_{add}(I,X)$, such that in the end 
 the complete fulvene Hamiltonian is written as in Eq.\,\ref{trampi}.

We consider a finite chain of six sites, whose Hamiltonian is
$H^{}=\sum |n \rangle \langle n+1 |+\text{H.c.}$ with $n=1,\cdots,5$
(hopping energy is set to unity)
depicted by straight lines in Fig.~\ref{fulvena}\,(a).
This chain is bipartite with $A$ points in the set $\{2, 4, 6 \}$ and $B$ points in the set $\{1, 3, 5 \}$.
$H$ is non-singular and satisfies the criteria leading to Eq.\,\ref{p11fig}.
We choose $A$ as the interference set, marked with $I$ in the \ref{fulvena}\,(a). Consequently, the set $B$ contains the rigid points, marked with $R$. To transform the linear chain into the fulvene molecule, a new term that links the $I$ points $2$ and $6$ is added to the Hamiltonian,
 $H_{add}^{}=|2 \rangle \langle 6 |+ \text{H.c.}$.

Consequently, the fulvene lattice
has the interference set ${\mathcal M}_{I}^{(a)}=\{2, 4, 6 \}$,
the rigid set ${\mathcal M}_{R}^{(a)}=\{1,3,5 \}$,
 while  the external set from Eq.\,\ref{desc2} is empty, ${\mathcal M}_{X}^{(a)}=\varnothing$.
 Figs.\,\ref{fulvena}\,(b) and (c) are variants of this scenario without any  $X$ points added.

Different decompositions of the total Hamiltonian is presented in Figures\,\ref{fulvena}\,(d), (e) and (f) where external points are considered.
In Fig.\,\ref{fulvena}\,(d),
 the bipartite system is chosen as $H=\sum |n \rangle \langle n+1 |+\text{H.c.}$ with $n=1,\cdots,3$, with
$\{2, 4 \}$ as $A$ points and $\{1, 3 \}$ as $B$ points.
In this case, $H$ is not singular and Eq.\,\ref{p11fig} is applied. Therefore we choose the $I$ points to be $A$
and $R$ points to be $B$, marked by $I$ and $R$ circles.
To recover the full fulvene lattice, two fresh points $X=\{5, 6 \}$ are added,
with new hopping energies represented by dashed lines.
$H_{add}$ contains terms that relate the $I$ points to $X$ points, $|2 \rangle \langle 6 |+ |4 \rangle \langle 5 | + \text{H.c.}$,
as well as the hopping between the two $X$ points, $|5 \rangle \langle 6 |+\text{H.c.}$.
The $H_{add}$ is of the type $H_2$ in Eq.\,\ref{trampi}.
With this, following the previous results, we obtain that the interference set is ${\mathcal M}_{I}^{(d)}=\{2, 4 \}$, ${\mathcal M}_{R}^{(d)}=\{1, 3 \}$, and ${\mathcal M}_{X}^{(d)}=\{5, 6 \}$.
Figs.\,\ref{fulvena}\,(e) and (f) represent also new decompositions having $X$ points added.

In Table\,\ref{tabelfulvena} we give the full list of the six sets of the special points
$I$, $R$ and $X$ obtained from the fulvene lattice.
It encodes all the $E = 0$ DQI processes that exist in the lattice
and the corresponding figures explain how they originate from a destructive interference in a bipartite system.

\begin{table}[h]
\caption{Fulvene special points: interference  points $I$, rigid points $R$ and external points $X$.
$I\in {\mathcal M}_{I}$, $R\in {\mathcal M}_{R}$ and $X\in {\mathcal M}_{X}$.
The robustness of every interference set ${\mathcal M}_{I}$ is dictated by the invariance set
${\mathcal M}_\text {inv}={\mathcal M}_{I} \cup {\mathcal M}_{X}$.} 
\label{tabelfulvena}
\centering
\begin{tabular}{|c|c|c|c|}
\hline
  Fig.\,\ref{fulvena} &      Interference  &  Rigid   &  External    \\
         &  points $I$      &     points $R$     &   points $X$   \\
\hline
  (a) &2, 4, 6  &  1, 3, 5  &     \\
\hline
  (b) &  2, 3, 5  &  1, 4, 6  &     \\
\hline
  (c) &  2, 4, 5  &  1, 3, 6  &     \\
\hline
  (d) &  2, 4  &  1, 3  & 5, 6    \\
\hline
  (e) &  2, 5  &  1, 6  & 3, 4     \\
\hline
  (f) &  2  &  1  &  3, 4, 5, 6    \\
\hline
\end{tabular}
%
\end{table}
\begin{table}[h]
\caption{The twelve DQI processes at $E=0$ for the fulvene lattice and
their invariance sets.}
\label{tabelzf}
\centering
\begin{tabular}{|c|c|}
\hline
  Green's function   &  Invariance                                     \\
  zeros &    at ${\mathcal M}_\text {inv}$                            \\
\hline
 $G_{22}, G_{23}, G_{24},$   & ${\mathcal M}_\text {inv}^{(f)}=\{ {2}, {3}, {4}, {5}, {6} \}$  \\
 $G_{25}, G_{26}$            &    \\
\hline
 $G_{44}, G_{46}$  & ${\mathcal M}_\text {inv}^{(d)}=\{ 2, 4, 5, 6 \}$   \\
\hline
 $G_{55}, G_{35}$  & ${\mathcal M}_\text {inv}^{(e)}=\{ 2, 3, 4, 5 \}$   \\
\hline
 $G_{33}$  & ${\mathcal M}_\text {inv}^{(b)}=\{ 2, 3, 5 \}$  \\
\hline
 $G_{66}$  & ${\mathcal M}_\text {inv}^{(a)}=\{ 2, 4, 6 \}$  \\
\hline
 $G_{45}$  & ${\mathcal M}_\text {inv}^{(d)}=\{ 2, 4, 5, 6 \}$   \\
           & ${\mathcal M}_\text {inv}^{(e)}=\{ 2, 3, 4, 5 \}$   \\
\hline
\end{tabular}
\end{table}

Different choices of the initial set of interference points lead to different predictions
of zeros of conductances
and to various invariance sets. 
It is therefore important that one finds all the possible zeros together
with there invariance sets, as summarized in Table\,\ref{tabelzf}.

We remark that when compared with all the other conductance zeros that are associated with a single invariance set, the $G_{45}$ zero is associated with  two.
From Figs.\,\ref{fulvena}\,(c), (d) and (e) or from the corresponding
lines of the Table\,\ref{tabelfulvena}, we note that
$G_{45}$ is a $G_{II}$ zero in (c) and a $G_{IX}$ zero in (d) and (e).
The three invariance sets of $G_{45}$ zero
are,
${\cal M}^{(c)}_\text {inv}= \{2, 4, 5\}$,
${\cal M}^{(d)}_\text {inv}= \{2, 4, 5, 6\}$ and
${\cal M}^{(e)}_\text {inv}= \{2, 3, 4, 5\}$.
The set ${\cal M}^{(c)}_\text {inv}$ can be ignored
 as it is included in the others two sets and one retains only two invariance sets,
${\cal M}^{(d)}_\text {inv}$ and ${\cal M}^{(e)}_\text {inv}$.

Another matter of interest is to say not only which perturbations keep the conductance zeros invariant
but also which ones lift them leading to nonzero current through a given molecular device
\cite{baer2002, shuguang2018, cardamone2006}.
Since we have shown that the conductance zeros are invariant under the $I$ and $X$ sites perturbations
it can be assumed that they may be destroyed by means of $R$ sites.
This however has to be investigated for every particular case.

\red{It is straightforward to show that in the case of a small molecule like fulvene,
the conductance zeroes can be individually predicted by alternative approaches,
such as the four graphs nullities investigation \cite{fowler2009}, determinant algorithm \cite{markussen2010, evers2020}
or by direct calculation from the power series expansion of the Green's function \cite{tsuji2018}.
We believe that our method becomes more practical when large molecules or large lattices are involved. Then,
once at least one bipartite sub-system is found, conductance zeroes can be predicted in group rather than individually, along with their robustness under perturbations.}

%


\section{Conclusions}

In this paper, we develop a method for determining the conductance zeros that result from destructive electron state interference in complex systems. The method starts by finding a smaller system which includes an interference set,
\red {whose points are such that any pair of them is} associated with zero conductance (${\bf G}_{II}=0$, \red{including the diagonal terms, when both leads are connected to the same site}).
The remaining points from this initial system,
outside the interference set, are called rigid points $R$. A complex system is then obtained by connecting new, external points $X$ to the $I$ points \red {(but never to the $R$ points)} or by adding any on-site or hopping energies related to the $I$ points.
We prove that the new system obtained in this way retains all the initial conductance cancellations between the $I$ points (${\bf G}_{II}=0$)
and, in addition, exhibits new zeros between points $I$ and points $X$ (${\bf G}_{IX}=0$).

Bipartite systems are known to have such sets of interference points at zero energy, provided that
$\epsilon =0$ is not an eigenvalue of the system \red{(i.e. Hamiltonian is non-singular)}, and they are used as starting blocks in our derivations. In such a case, the sets of $I$ points can be identified with the sub-lattices points A or B. Among the simplest examples of bipartite systems with non-singular Hamiltonians we mention the linear chains with $2N$ sites and circular molecules with $4N+2$ sites.


The Dyson expansion \blue{is used to prove the conductance cancelations as well as their invariance properties. The zeros are robust against perturbations
applied on the $I$ and $X$ points, while the $R$ points should not be perturbed.}


In the case of the non-bipartite fulvene molecule it is shown that, by choosing in different ways the initial set of
$I$ points, one can obtain all the conductance zeros and study their persistence under the effect of different perturbations.

Our study contributes to the understanding
of the destructive interferences and their invariance properties
 for an appropriate class of physical systems that have bipartite lattices or contain subsystems with bipartite lattices.
 They can be relevant for
transport experiments on molecules, nanostructures and various finite lattices,
for designing of logical gates or for projection of
quantum interference transistors at the nanoscale.

\acknowledgements
The work is supported by Romanian Core Program PN19-03 (contract no.\,21\,N/08.\,02.\,2019).
The authors thank professor Alexandru Aldea and Bogdan Ostahie for useful discussions and valuable help.


\appendix

\section{Conductance Calculation}\label{conductance}

In this Section we briefly remind how the Green's function cancellations reflect also in the conductance, should the system no longer be isolated, but connected to external leads.
Let us consider for this a two-terminal conductor, such as a tight-binding lattice
connected at two semi-infinite leads called $L_{out}$ and $L_{in}$.
They are coupled at two sites of the involved lattice $i$ and $j$, respectively.
Transport coefficients are given in terms of the effective Green's function
of the Hamiltonian $H^\text{eff}=H+\tau_i |i \rangle \langle i | + \tau_j |j \rangle \langle j |$.
This is evaluated as in Refs.~\citenum{ostahie2016,tolea2010} for a wave vector $k$ and Fermi energy  $E=2\tau_l\cos k$ as a function of $\tau_l$, the hopping energy on the leads, and $\tau_c$, the hopping energy between the leads and the discrete system. Therefore,
\begin{eqnarray}\label{geff}
&&{ G}^\text{eff}_{ij}(E)=\frac{{ G}_{ij}(E)}
{1-\tau_j { G}_{jj}-\tau_i { G}_{ii}
-\tau_j\tau_i{ G}_{ij}{ G}_{ji}+\tau_j\tau_i{ G}_{ii}{ G}_{jj}}\;,\nonumber\\
&& \text{with}~\tau_{i,j}=\frac{\tau_c^2}{\tau_l}e^{-ik}.
\end{eqnarray}
%

In the Landauer-B\"uttiker approach \cite{landauer1957, buttiker1986, imry1999}
the tunneling amplitude from the lead $L_{in}$ into
the lead $L_{out}$ gives the scattered wave function in the lead $L_{out}$,
$|\Psi_{L_{out}}\rangle={\bf t}_{out,in} \sum_{n_l\in L_{out}} e^{-ikn_l}|n_l\rangle$.
The tunnelig amplitude from $L_{in}$ to $L_{out}$ is
\begin{eqnarray}\label{tun}
{\bf t}_{out,in}(E)=2 i \frac{\tau_c^2}{\tau_l}  \sin k \;{ G}^\text{eff}_{ij}(E)\,. 
\end{eqnarray}
The electric conductance ${\bf G}_{ij}$ or transmittance ${\bf T}_{ij}$ are given:
\begin{eqnarray}\label{tra}
{\bf G}_{ij}(E)=\frac{e^2}{h}{\bf T}_{ij}(E)
=\frac{e^2}{h}|{\bf t}_{out,in}(E)|^2\,. 
\end{eqnarray}

Straightforwardly, the effective Green's function
${ G}^{\text{eff}}_{ij}(E)$ and of course the conductance ${\bf G}_{ij}(E)$ cancels
whenever the Green's function
of the isolated sample ${ G}_{ij}(E)$ is equal to zero.

\blue{We also point out again that the Green's function cancellations obtained in Eqs.\,\ref{gprim} and \ref{gprim2}
are proved for
nonsingular $H'$ such that $G'$ Green's functions in Dyson expansions from Eqs.\,\ref{gprim0} and  \ref{gprimext}
have no singularities at energy $E$.
Otherwise, supplementary calculations have to be carried out, by using for instance effective Hamiltonian
depicted here, to directly prove the conductance
cancellations. With some exception (as the square Hamiltonian in \cite{sykora2017b})
$H^{\text {eff}}$ is nonsingular because of the non-hermitian terms added to the contact points.
}

\newpage
~~

\newpage

\end{document}